\documentclass[11pt]{article}
\usepackage{jheppub}
\usepackage{braket} 

\def\tr{{\rm tr}}

\def\CF{{\cal F}}

\def\CH{{\cal H}}

\def\CR{{\cal R}}
\def\CM{{\cal M}}

\def\CO{{\cal O}}

\def\BH{\mathbb{H}}
\def\BR{\mathbb{R}}
\def\BS{\mathbb{S}}
\def\BT{\mathbb{T}}

\title{Renormalized Entanglement Entropy on Cylinder}

\author[a]{Shamik Banerjee,}
\author[a,b]{Yuki Nakaguchi,}
\author[b]{and Tatsuma Nishioka}

\affiliation[a]{Kavli Institute for the Physics and Mathematics of the Universe, The University of Tokyo,\\
5-1-5 Kashiwa-no-Ha, Kashiwa City, Chiba 277-8568, Japan}
\affiliation[b]{Department of Physics, Faculty of Science,
The University of Tokyo,\\
Bunkyo-ku, Tokyo 113-0033, Japan}

\emailAdd{banerjeeshamik.phy@gmail.com}
\emailAdd{yuuki.nakaguchi@ipmu.jp}
\emailAdd{nishioka@hep-th.phys.s.u-tokyo.ac.jp}
                
\abstract{
We develop a framework of calculating entanglement entropy for non-conformal field theories with the use of the dilaton effective action.
To illustrate it, we locate a theory on a cylinder $\BR \times \BS^{2}$ and compute entanglement entropy of a cap-like region perturbatively with respect to the mass for a free massive scalar field.
A renormalized entanglement entropy (REE) is proposed to regularize the ultraviolet divergence on the cylinder.
We find that the REE decreases monotonically both in the small and large mass regions as the mass increases.
We confirm all of these behaviors by the numerical calculations, which further shows the monotonic decrease of the REE in the entire renormalization group flow.
}

\preprint{IPMU-15-0124, UT-15-28}

\begin{document}

\maketitle

\section{Introduction}
Quantum aspects of field theories manifest themselves in various forms including anomalies, renormalization group (RG) flow, and strong dynamics on the long distance.
Some of them can be uncovered through correlation functions of local operators that are only sensitive to the short-scale structure, while non-local observables like Wilson loops are needed to detect long-range correlations in quantum field theories (QFTs).
Among such non-local quantities is entanglement entropy (EE) that diagnoses quantum entanglement relevant at low energy where thermal (classical) fluctuations are suppressed enough.
In critical phenomena, EE shows universal and characteristic features that reflect the central charges of conformal field theories \cite{Holzhey:1994we,Vidal:2002rm,Calabrese:2004eu,Ryu:2006bv,Ryu:2006ef}  and the degeneracy of the ground state of topological QFTs \cite{Kitaev:2005dm,levin2006detecting,Grover:2011fa}.
It also plays an important role as a measure of degrees of freedom under RG flows and confinement/deconfinement transitions \cite{Casini:2004bw,Liu:2012eea,Casini:2012ei,Ryu:2006bv,Ryu:2006ef,Nishioka:2006gr,Klebanov:2007ws,Pakman:2008ui,Myers:2010xs,Myers:2010tj}.

Let us assume that the total system is in its ground state $|\Omega\rangle$. 
We denote the subsystem for which we want to compute the EE by $A$. 
The entangling surface is the boundary of $A$ which we denote by $\Sigma \equiv \partial A = \partial \bar A$. The density matrix describing the total system is given by
\begin{equation}
\rho_\text{vac} = |\Omega\rangle \langle\Omega| \ .
\end{equation}
If the Hilbert space of the theory is assumed to factorize as $\CH= \CH_{A}\otimes \CH_{\bar A}$, one can define the reduced density matrix of $A$ by
\begin{equation}
\rho_{A} = \tr_{\CH_{\bar A}}\, \rho_\text{vac} \ .
\end{equation}
The EE of the subsystem $A$ is then defined as the Von Neumann entropy associated with the reduced density matrix $\rho_{A}$,
\begin{equation}
S_A = - \tr_{\CH_{A}} \, \rho_{A} \ln \rho_{A}\ .
\end{equation} 

In QFTs, EE is usually calculated by the replica trick.
Applying the replica trick, EE can be computed using the formula \cite{Holzhey:1994we,Calabrese:2004eu}
\begin{equation}\label{ReplicaTrick}
S_A = \lim_{n\to 1} \partial_n (F_{n} - nF_{1})\ .
\end{equation}
Here $F_n \equiv -\ln Z[\CM_n]$ is the free energy of a theory on the $n$-fold cover $\CM_n$ of the Euclidean space-time on which the theory lives. 
Let us describe this in some detail.

The entangling surface $\Sigma$ is a codimension-two hypersurface in the (Euclidean) space-time.
Parameterizing the transverse two-dimensional space by the polar coordinates $(\rho,\tau)$ so that $\Sigma = \{ \,\vec x \,| \,\rho=0\}$,
the $n$-fold cover $\CM_n$ is obtained by extending the periodicity of the angular coordinate $\tau$ from $2\pi$ to $2\pi n$. 
For $n\neq 1$, the $n$-fold cover $\CM_n$ has a conical singularity at $\Sigma$ of the angular excess $2\pi(n-1)$.
This singularity prevents us from calculating the exact free energy $F_n$ of QFTs except for free field theories \cite{Kabat:1995eq,Klebanov:2011uf}.
In practice it is not easy to work on such singular spaces in particular when the translational symmetry along $\tau$ is broken in the transverse space. 
A common practice is to regulate the conical singularity by replacing the conifold with a smooth manifold with a regulator parameter and remove the regulator at the end of the calculation. 
This procedure is ambiguous because there are more than one methods of regularization and it is not clear whether all of them give the same answer for EE. Instead one could work on the singular conifold without regularization.

It is, however, not so obvious how to implement conformal perturbation theory on a cone. 
Proper determination of boundary conditions at the singularity is difficult. 
In this paper we will attack this problem in a different way. 
We will call it Conformal Decompactification. The idea is to perform a conformal transformation of the replica space such that the resulting space has no conical singularity and calculate the free energy on the non-singular geometry. The only constraint on the conformal factor is that it should blow up along the entangling surface where the conical singularity is located. For non-conformal field theories the two free energies are made equal by switching on a background dilaton field on the non-singular space.

In section \ref{ss:ConfDecom}, we will describe the conformal decompactification in a relevant perturbation of conformal field theories (CFTs) 
and provide a framework for the perturbative calculation of the EE. 
In general, the perturbation theory of EE on non-compact spaces is plagued by infrared (IR) divergences.
In order to avoid such IR divergences, we will locate the theory on a compact space such as a cylinder $\BR\times \BS^{d-1}$ with the metric
\begin{align}\label{CylinderMetric}
        ds^2 = -dt^2 + R^2\left(d\theta^2 + \sin^2\theta\, d\Omega_{d-2}^2 \right) \ ,
\end{align}
and consider the EE of a cap-like region $A = \{\,\vec x\, |\, t=0,\, 0\le \theta \le \theta_0 \}$.
This type of EE has been extensively studied in recent literatures \cite{Herzog:2014fra,Sabella-Garnier:2014fda,Herzog:2014tfa,Herzog:2015cxa}, but
it remains unclear how to calculate the entropy and how it behaves away from conformal fixed points.
We wish to explore aspects of the EE on the cylinder \eqref{CylinderMetric} in non-CFTs.


For brevity, we will confine ourselves to the case in $d=3$ dimensions, where the entropy on the cylinder \eqref{CylinderMetric} has the area law divergence proportional to $R\sin\theta_0$ that needs to be regularized to extract a scheme-independent part from EE.
The situation is the same for EE of a disk of radius $R_\text{disk}$ on a flat space and 
we briefly review the known facts so as to extend them to the case we are dealing with.

A simple and pragmatic regularization of the UV divergence of EE for a disk is proposed by Liu and Mezei \cite{Liu:2012eea} as the renormalized entanglement entropy (REE):
\begin{align}\label{LMREEoriginal}
        \CF(R_\text{disk}) = (R_\text{disk}\partial_{R_\text{disk}} - 1)S_A(R_\text{disk}) \ .
\end{align}
Among the other various regularizations, the REE turns out to have particularly important properties as a measure of degrees of freedom in QFT:
\begin{enumerate}
\item It is monotonic $\CF' \le 0$ and the equality holds only for a theory being conformal \cite{Casini:2012ei}.
\item The value of $\CF$ for CFT$_3$ is the same as the free energy on a three-sphere, $F\equiv -\log Z_{\BS^3}$ \cite{Casini:2011kv}.
\end{enumerate}
Upon the identification of the size of the disk with the inverse of the renormalization scale, the REE monotonically decreases under any RG flow, proving the $F$-theorem that states $F_\text{UV}\ge F_\text{IR}$ for the free energies at the UV and IR conformal fixed points \cite{Jafferis:2011zi,Klebanov:2011gs,Myers:2010xs,Myers:2010tj}.
We will call a function satisfying these properties an $F$-function.
An $F$-function is reminiscent of the Zamolodchikov's $c$-function \cite{Zamolodchikov:1986gt} that also satisfies stationarity in addition to the counterparts of the two properties.\footnote{The second property of the Zamolodchikov's $c$-function is that the value at the fixed points is the same as the central charge of the CFT$_2$.}
In our case, the stationarity of an $F$-function can be stated in the following way:
\begin{enumerate}
\item[3.] The derivative of $\CF$ with respect to a coupling constant $g$ of a relevant operator vanishes at the conformal fixed points of RG flows, $\partial_g \CF |_\text{CFT}= 0$.
\end{enumerate}
The third property, not necessary for the proof of the $F$-theorem, can be favorable for finding the fixed points of RG via an $F$-function.
The stationarity of the REE \eqref{LMREEoriginal}, however, is doubted for a massive free scalar theory based on the numerical calculation \cite{Klebanov:2012va,Nishioka:2014kpa} where it is speculated that the non-stationarity of the REE originates from the IR divergence of EE.

One of the aims of this paper is to seek an $F$-function in the Zamolodchikov's sense, i.e.\ , satisfying all the aforementioned three properties, through the EE on a cylinder which is free from the IR divergence.

There are two natural generalizations of the REE on the cylinder as will be introduced in section \ref{ss:REECylinder}.
One is the Liu-Mezei type, called $\CF_\text{LM}$, which takes the same form as \eqref{LMREEoriginal} with $R_\text{disk}$ replaced with the size of the sphere $R$ while fixing the angle $\theta_0$.
On the other hand, one can define another function, called $\CF_\text{C}$, by replacing $R_\text{disk}$ with $\sin\theta_0$ in \eqref{LMREEoriginal} while fixing $R$, to remove the UV divergence.
We will show both the REEs on the cylinder are finite, coincide with the three-sphere free energy $F$ at RG fixed points, and reduce to the original REE \eqref{LMREEoriginal} in the flat space limit $R\to\infty$ and $\theta_0\to 0$ while keeping $R\theta_0 \equiv R_\text{disk}$ held fixed.
The REEs are weak $F$-functions in a sense that their values at the UV fixed point are larger or equal to those at the IR fixed point as a consequence of the $F$-theorem.

Reminding the situation for the REE on the flat space, a free massive scalar field is a good testing ground for examining the stationarity of the REEs on the cylinder.
As the entropy is dimensionless, $S_A$ is a function of dimensionless parameters $\theta_0$ and $mR$ for a  scalar field of mass $m$.
In section \ref{ss:FreeMassiveScalar}, we carry out the perturbative expansions of $S_A(\theta_0, mR)$ in the small and large $mR$ limits respectively. 
In the small $mR$ limit, we apply the conformal decompactification developed in section \ref{ss:ConfDecom} to the cylinder and retrieve the leading contribution of order $(mR)^2$ to the entropy. 
In the other limit, we slightly generalize the method of \cite{Klebanov:2012yf} that relates the order $1/(mR)$ term in the large $mR$ expansion with the logarithmic divergence in the EE of a four-dimensional free massless scalar field.
In both regimes, we will show both of our REEs, $\CF_\text{LM}$ and $\CF_\text{C}$, decrease monotonically as $mR$ becomes large for fixed $\theta_0$ and as $\theta_0$ increases from $0$ to $\pi/2$ for fixed $mR$.
Meanwhile, we find $\CF_\text{C}$ is stationary at the UV fixed point ($mR=0$), being a good candidate for an $F$-function, while $\CF_\text{LM}$ still is not stationary even on the cylinder.

The limiting behaviors of $\CF_\text{C}$ are confirmed by the numerical calculation which further demonstrates the monotonicity under the entire RG flow.
The numerical calculation is based on the real time approach \cite{Bombelli:1986rw,Srednicki:1993im} which does not rely on the replica trick.
Thus it supports the validities of the conformal decompactification and the large mass expansion we use in our analytic calculations.

Put it all together, the examination of the EE of a free massive scalar field provides us a non-trivial evidence for $\CF_\text{C}$ being a strong $F$-function that monotonically decreases under any RG flow.
It satisfies the three properties at least for a free massive scalar field, and could be an $F$-function in the Zamolodchikov's sense.
It would be intriguing to extend the proof of the $F$-theorem \cite{Casini:2012ei,Casini:2015woa} to the cylinder, which might prove the monotonicity of our REEs.
On the other hand, we were not able to verify the monotonicity of $\CF_\text{LM}$ even numerically.
Moreover, we are led to a surprising discrepancy between the numerical result and the analytic large mass expansion that arises only for $\CF_\text{LM}$.
We will give the details of the numerical algorithm and the conjectured form of the large mass expansion fitted from the result in section \ref{ss:discussion} where possible sources of the inconsistency are also discussed.


\section{Conformal decompactification}\label{ss:ConfDecom}

We illustrate the conformal decompactification by taking our space-time to be $(2+1)$-dimensional Minkowski space and the subsystem to be a disk of radius $R$ on some spatial plane. The entangling surface is a circle of radius $R$. Since we are interested in a static situation we can as well do everything in the Euclidean space $\BR^3$. In a coordinate system adapted to the disk the metric of $\BR^3$ can be written as
\begin{equation}\label{R3metric}
ds^2_{\BR^3}= d\rho^2 + \rho^2 d\tau^2 +(R+\rho \,\cos\tau)^2 d\phi^2 \ .
\end{equation}
The coordinates $\rho$ and $\tau$ are the radial and angular coordinates, respectively, on the plane transverse to the circular entangling surface located at $\rho=0$. 
The angular coordinate $\phi$ along the entangling circle has periodicity $2\pi$. 

The replica trick amounts to changing the periodicity of $\tau$ from $2\pi$ to $2\pi n$, where $n$ is a positive integer. If $n\neq 1$, then there is a conical singularity at $\rho=0$ with the angular excess $2\pi(n-1)$. 
In this example, we have a conical singularity because the transverse $\tau$-circle is shrinking to zero size along the entangling surface. One way of getting a regular space is to perform a conformal transformation which blows up along the entangling surface. In the resulting conformally transformed space the transverse circle will be non-contractible with coordinate periodicity $2\pi n$ and there will be no conical singularity. We can read off the free energy on the conifold from the free energy on the conformally related smooth manifold if we know how the free energies are related.\footnote{The reader should note that this method is similar in spirit to the method of \cite{Lewkowycz:2013nqa}.}

Let us see in the particular case of the disk how this conformal transformation can be done. Following \cite{Casini:2011kv,Hung:2014npa} we write the metric of $\BR^3$ in the cylindrical polar coordinates as
\begin{equation}
ds^2_{\BR^3} = dt_E^2 + dr^2 + r^2 d\phi^2 \ ,
\end{equation}
where $t_E$ is the Euclidean time and the entangling circle is located at $(t_E,r)=(0,R)$. This is a different coordinate system from that was used in \eqref{R3metric}. 
If we now define
\begin{equation}
 \omega = r+ it_E\ , \qquad \Sigma= u+ i\tau_E \ ,
\end{equation}
and make the following coordinate transformation 
\begin{equation}
 e^{-\Sigma} = \frac{R-\omega}{R+\omega} \ ,
\end{equation}
the metric of $\BR^3$ can be written as
\begin{equation}
R^2(d\tau_E^2 + du^2 + \sinh^2 u \ d\phi^2) = e^{2\sigma} ds_{\BR^3}^2 \ ,
\end{equation} 
where the conformal factor is given by
\begin{equation}
e^{\sigma} = \frac{2R^2}{|R^2 - \omega^2|} \ .
\end{equation}

The metric on the left hand side is the metric of two-dimensional hyperbolic space $\BH^2$ times a circle parameterized by $\tau_E$ with periodicity $2\pi$. 
If we extend the periodicity of $\tau_E$ from $2\pi$ to $2\pi n$, then the metric of $\BH^2\times \BS^1$ becomes conformal to the replica geometry where there is a conical singularity along the entangling circle located at $\omega=R$ of angular excess $2\pi (n-1)$. 
The conformal factor is independent of $n$ and is a periodic function of $\tau_E$ with periodicity $2\pi$. 
Hence the replica geometry corresponding to a disk can be conformally mapped to $\BH^2\times \BS^1$ with $\BS^1$ periodicity $2\pi n$ and the conformal factor blows up along the entangling circle at $\omega=R$. 
The $\BS^1$ factor is non-contractible in the resulting geometry and this $\BS^1$ is precisely the image of the contractible circle in the plane transverse to the entangling surface located at $\omega=R$, which is used to perform the replica trick. 
 
Although we have explained the procedure for a disk, this method is general because every entangling surface is locally the same. There is a contractible circle in the plane transverse to the entangling surface and we can make it non-contractible by doing a conformal transformation which blows up along the entangling surface. The resulting geometry will have a non-contractible circle and will be smooth. For example, if we take as our subsystem a two-dimensional region bounded by the curve, $r=f(\phi)$, where $f(\phi)$ is a periodic single-valued function of $\phi$, then we can choose the conformal factor to be of the form: 
\begin{equation}
 e^{\sigma} = \frac{2f^2(\phi)}{|f^2(\phi) - \omega^2|} \ .
\end{equation}
The coordinates $\phi$ and $\omega$ are as defined before. The conformal factor diverges along the curve $r=f(\phi)$ located on the $t_E=0$ plane. The resulting geometry obtained by multiplying the $\BR^3$ metric by this conformal factor is no longer $\BH^2\times \BS^1$ but it is smooth and has a non-contractible circle. This circle has periodicity $2\pi$ and every field theory quantity is periodic under $2\pi$ shift along the circle. The replica trick in this geometry amounts to making the periodicity of the circle $2\pi n$. This does not produce any singularity because $n$ is a positive integer and things are periodic along the circle with periodicity $2\pi$.

\subsection{Relating the free energies on conformally related spaces} 

For a conformal field theory the free energies on the replica space and the conformally transformed space are the same modulo the conformal anomaly if the space-time dimension is even. The conformal anomaly part can be computed by standard methods. 

For a general non-conformal field theory the free energies on the two spaces are not equal, but they can be related if we introduce a background dilaton field, which we denote by $\tau(x)$. Let us denote the metric of the replica space by $g_{\mu\nu}(x)$.

If $Z[g_{\mu\nu}(x),\tau(x)]$ is the partition function of the Euclidean theory in the presence of the background metric $g_{\mu\nu}(x)$ and dilaton field $\tau(x)$, then it satisfies the following transformation rule \cite{Komargodski:2011vj, Komargodski:2011xv, Luty:2012ww, Baume:2014rla}
\begin{equation}
Z[e^{2\sigma(x)}g_{\mu\nu}(x), \tau(x) + \sigma(x)] = C\, Z[g_{\mu\nu} (x),\tau(x)] \ ,
\end{equation}
where $C$ is completely determined by the conformal anomaly of the ultraviolet (UV) CFT and does not depend on the mass parameters of the theory. In particular $C=1$ in odd dimensions due to the absence of conformal anomaly. 
The free energy defined as $F=- \ln Z$ satisfies the relation:
\begin{equation}\label{dileqa}
F[e^{2\sigma(x)}g_{\mu\nu}, \tau(x) + \sigma(x)] =  F[g_{\mu\nu},\tau(x)] \ ,
\end{equation}
where we have neglected $\ln C$, because we are only interested in the part of the EE generated by the massive deformation. 
In odd dimensions this factor is identically zero and this equality is exact. In even dimensions this anomaly part gives rise to local terms in the dilaton effective action some of which are uniquely determined by the trace anomaly matching. These local terms in the dilaton effective action give the logarithmically divergent universal terms in the entanglement entropy which were computed by using this technique in  \cite{Banerjee:2014daa, Banerjee:2014hqa}.

Now the equality \eqref{dileqa} holds for any functional form of the dilaton field $\tau(x)$ and we can also write
\begin{equation}\label{FreeEnergyRelation}
F[e^{2\sigma(x)}g_{\mu\nu}, \sigma(x)] =  F[g_{\mu\nu},\tau(x)=0] \ .
\end{equation}
The right hand side represents the free energy on the replica space in the absence of the dilaton field, which is precisely what we want to compute, and the left hand side represents the free energy on the conformally related non-singular space but in the presence of a background dilaton field which is equal to the conformal factor $\sigma(x)$. We will use this relation to compute the EE by conformally mapping the problem to a non-singular space.

\subsection{Deformed CFT coupled to dilaton}
Let us consider a UV CFT in $d$ dimensions deformed by some (marginally) relevant operator $\CO_{\Delta}$ of dimension $\Delta$. 
The action can be written as
\begin{equation}
I = I_\text{UV\,CFT} + \mu^{d-\Delta} \int d^{d}x \,\sqrt{g} \, \lambda(\mu) \, \CO_{\Delta} \ ,
\end{equation}
where $\lambda(\mu)$ is the dimensionless renormalized coupling constant at renormalization scale $\mu$. 
It is determined by the beta function equation
\begin{equation}
\mu \frac{d}{d\mu} \lambda = \beta (\lambda) \ .
\end{equation}

After conformally transforming to the non-singular space the free energy has to be calculated on the new geometry with dilaton turned on. 
The dilaton field couples to the field theory as \cite{Komargodski:2011xv}\footnote{See appendix \ref{ss:appA} for an example of the action on a conformally transformed manifold.}
\begin{equation}
\tilde I = \tilde I_\text{UV\,CFT} + \mu^{d-\Delta} \int d^{d} x \,\sqrt{\tilde g} \, \lambda(\mu e^{\tau})\, \CO_{\Delta} \ .
\end{equation}
If we set $\tau=\sigma$ where $g_{\mu\nu} = e^{-2\sigma} \tilde g_{\mu\nu}$, we get the following action
\begin{equation}
\tilde I = \tilde I_\text{UV\,CFT} + \mu^{d-\Delta}\int d^{d}x\, \sqrt{\tilde g} \, \lambda(\mu e^{\sigma}) \, \CO_{\Delta} \ ,
\end{equation}
which needs to be used for the calculation of free energy on the conformally transformed manifold. 
To summarize, in our prescription, one needs to calculate the dilaton (with $\tau=\sigma$) effective action on the conformally related non-singular manifold to compute the EE.

We would like to emphasize that we have not used any perturbation theory to arrive at this prescription. 
Thus it can be used even if conformal perturbation theory breaks down.

\section{Renormalized entanglement entropy on cylinder}\label{ss:REECylinder}

We will consider a theory on a cylinder $\BR \times \BS^{d-1}$ with the metric given by \eqref{CylinderMetric}
and divide the $\BS^{d-1}$ by a codimension-two hypersurface $\Sigma$ at $t=0$ and $\theta = \theta_0$ to a subsystem $A$ within $0\le \theta \le \theta_0$ and its compliment $\bar A$ within $\theta_0 \le \theta \le \pi$.
The angle $\theta_0$ can be restricted to be $0\le \theta_0\le \pi/2$ for the entropy is symmetric with respect to the exchange of $A$ and $\bar A$ when we concentrate only on the vacuum state of the theory.

Employing the replica trick, one can calculate the EE with the partition function on the $n$-fold cover of the Euclidean space of \eqref{CylinderMetric}
\begin{align}\label{S^1timesS^2}
        ds^2  = dt_E^2 + R^2\left( d\theta^2 + \sin^2\theta\, d\Omega_{d-2}^2\right) \ ,
\end{align}
that has a surplus angle $2\pi (n-1)$ around $\Sigma$.
To make it transparent, we use the coordinate transformation
\begin{align}
        \begin{aligned}
                \tanh (t_E/R) &= \frac{\sin\theta_0\, \sin\tau}{\cosh u + \cos\theta_0\,\cos\tau} \ , \\
                \tan \theta &= \frac{\sin\theta_0\, \sinh u}{\cos\theta_0 \,\cosh u +\cos\tau} \ ,
        \end{aligned}
\end{align}
with $0\le u <\infty$ and $0\le \tau \le 2\pi$ for $n=1$.
The resulting metric becomes \cite{Casini:2011kv}
\begin{align}\label{S^1timesH^2}
        \begin{aligned}
                e^{2\sigma} ds^2 =&\, \, R^2\left[ d\tau^2 + du^2 + \sinh^2 u \, d\Omega_{d-2}^2 \right] \ , \\
                \ e^{-2\sigma}&\equiv \frac{\sin^2 \theta_0}{(\cos\tau + \cos\theta_0\,\cosh u)^2 + \sin^2\theta_0 \sinh^2u } \ .
        \end{aligned}
\end{align}
The $n$-fold cover is given by the metric \eqref{S^1timesH^2} with the period $\tau \sim \tau + 2\pi n$, which is conformally equivalent to $\BS^1\times \BH^{d-1}$.
We will denote the conformally equivalent manifold as $\BS^1_n \times \BH^{d-1}$.

The entangling surface $\Sigma$ located at $t_E = 0$ and $\theta = \theta_0$ in the original coordinates \eqref{S^1timesS^2} is mapped to $u=\infty$ in the new coordinates \eqref{S^1timesH^2} where the conformal factor $e^{2\sigma}$ blows up.
Note that the $\BS^1$ factor along $\tau$ is non-contractible in the resulting geometry. 
This $\BS^1$ is the image of the contractible circle in the plane transverse to the entangling surface at $t_E = 0$ and $\theta = \theta_0$, which is used to perform the replica trick. 

EE is always accompanied by UV divergences in QFT. 
The leading part is well-known as the area law term diverging as $1/\epsilon^{d-2}$ in $d$ dimensions for the UV cutoff $\epsilon\ll 1$.
For this reason, the bare entropy is scheme-dependent and needs to be renormalized so as to be free from the UV divergences.
One possible regularization is to renormalize the divergences to parameters in the background gravity theory such as the Newton and cosmological constants as is usually done in QFTs on curved spaces \cite{Birrell:1982ix}.

A simpler regularization was proposed by Liu and Mezei \cite{Liu:2012eea} for a spherical or any scalable entangling region on a flat space-time.
They define the renormalized entanglement entropy (REE) by acting a differential operator of the radius of the sphere on the EE.
In three dimensions, the REE of a disk of radius $R_\text{disk}$ becomes \eqref{LMREEoriginal}
which subtracts the UV divergence of the EE. Moreover it has been shown that the REE defined in this way is monotonically decreasing along any RG flow in three dimensions \cite{Casini:2012ei}, known as the $F$-theorem \cite{Jafferis:2011zi,Klebanov:2011gs}.

In our case we can define two types of REEs on the cylinder. 
First we note that the finite part of the EE of the cap-like region $A$ on $\BS^2$ equals to that of a disk on $\BR^2$ if the theory is conformal.
The finite part of the EE is minus the finite part of the $\BS^3$ free energy, $F\equiv - \log Z_{\BS^3}$, as was shown by \cite{Casini:2011kv}.
Thus the entropy $S_A (\theta_0)$ for CFT$_3$ on the cylinder takes the form:
\begin{align}\label{EEonCylinder}
        S_A(\theta_0)|_\text{CFT} = \alpha \frac{2\pi R\sin\theta_0}{\epsilon} - F \ ,
\end{align}
with a non-universal coefficient $\alpha$.
The first term, fixed by requiring the area law, is proportional to the circumference $R\sin\theta_0$ of the entangling surface.

A straightforward generalization of \eqref{LMREEoriginal} is to define the REE on the cylinder as
\begin{equation}\label{RenEE1}
 \CF_\text{LM} (R,\theta_0) = (R\partial_R - 1)S_A(R) |_{\theta_0} \ ,
\end{equation}
where the derivative with respect to $R$ is taken at fixed angle $\theta_0$. 
It is finite and becomes $\CF_\text{LM} = F$ at any RG fixed point thanks to the relation \eqref{EEonCylinder}.
We will see in section \ref{ss:FreeMassiveScalar} that for a massive scalar field of mass $m$, the REE $\CF_\text{LM}(mR)$ is monotonically decreasing in the small and large $mR$ regions as $mR$ increases at fixed $\theta_0$. 
At $mR=0$ it takes the value $F_\text{scalar}$ for a scalar field in three dimensions and decreases to $0$ as $mR\rightarrow \infty$. 
We, however, were not able to determine the shape of $\CF_\text{LM}$ in the intermediate regime $1\ll mR \ll \infty$ even numerically because of the finite lattice size effect.
Thus we do not know if the $\CF_\text{LM}$-function monotonically decreases along the entire RG flow of a massive scalar field theory.
It is to be noted that this gives rise to a family of $F$-functions parametrised by $\theta_0$ and they all interpolate between the UV and the IR fixed points of a massive scalar field in three dimensions, but none of them are stationary\footnote{The REE for a relevant perturbation of CFT is called stationary if the first derivative with respect to the coupling constant vanishes at a conformal fixed point. The REE of a disk \eqref{LMREEoriginal} is known to be non-stationary \cite{Klebanov:2012va,Nishioka:2014kpa} for a massive free scalar theory.} as function of $(mR)^2$ at the UV fixed point as will be shown in the next section.

The second way to renormalize the UV divergence of the entropy is to define the REE on the cylinder as
\begin{align}\label{RenEE2}
        \CF_\text{C}(R,\theta_0) \equiv \left( \tan\theta_0\, \partial_{\theta_0} - 1\right) S_A (\theta_0) |_{R} \ ,
\end{align}
where the derivative with respect to $\theta_0$ is taken at fixed $R$. As the definition implies, $\CF_\text{C}$ is always finite for the differential operator kills the area law divergence.
Also it coincides with the finite part of the $\BS^3$ free energy $F$ at a conformal fixed point.  We will see in the next section for a free massive scalar field of mass $m$ that $\CF_\text{C}$ decreases monotonically as a function of $(mR)^2$ at fixed $\theta_0$ and it is also stationary as a function of $(mR)^2$ at the UV fixed point. 
Then the REE $\CF_\text{C}$, obtained from EE on the cylinder, serves as an $F$-function in three dimensions.
It decreases monotonically from the UV to the IR and is stationary at the UV fixed point for a massive scalar field. 
This is analogous to the Zamolodchikov's $c$-function in two dimensions, at least for a massive scalar field.

Before closing this section, we comment on the flat space limit of the cylinder EE.
The cylinder metric \eqref{S^1timesS^2} reduces to the flat space in the $R\to \infty$ and $\theta \to 0$ limits with $r\equiv R\,\theta$ held fixed, and
the cap-like entangling region $A$ turns into the disk of radius $R_\text{disk} \equiv R\,\theta_0$.
It follows from their definitions that the two REEs \eqref{RenEE1} and \eqref{RenEE2} lead to the REE of a disk \eqref{LMREEoriginal} in this limit.

\section{Free massive scalar field}\label{ss:FreeMassiveScalar}
We will calculate the EE of the cap-like region $A$ on the cylinder for a free massive scalar field.
We assume that the scalar field is conformally coupled to the background geometry in the massless limit, whose action takes the form of
\begin{align}\label{ScalarAction}
        I = \frac{1}{2} \int d^{3}x\, \sqrt{g} \  \left[ g^{\mu\nu}\partial_\mu\phi \partial_\nu \phi + \frac{\CR}{8}\phi^{2} +  m^2\phi^{2}\right] \ ,
\end{align}
where $\CR$ is the Ricci scalar.
Applying the conformal decompactification and regarding the theory as a relevant perturbation of a free massless scalar theory by the mass term, the entropy will be expanded in the small mass limit and the leading term of order $m^2$ will be evaluated.
On the other hand, the large mass expansion will be carried out following \cite{Klebanov:2012yf} and the order $1/m$ term of the entropy will be fixed for a general entangling surface.
Finally the results in the two limits will be confirmed by the numerical calculation that shows the REE, $\CF_\text{C}(\theta_0, mR)$, monotonically decreases as $mR$ becomes large.
We also comment on the properties of the other REE, $\CF_\text{LM}$, and the obstacles we encounter in calculating it numerically on lattice.

\subsection{Small mass expansion}
The EE is expected to have a series expansion with respect to the scalar mass in the small mass region.
In order to fix the leading term of the expansion we are to calculate the derivative of the free energy $F_{n}$ on the $n$-fold cover $\CM_n$ of $\BR\times \BS^2$
\begin{equation}\label{dFndm2}
\frac{\partial}{\partial m^{2}} F_{n} = \frac{1}{2} \int_{\CM_n} d^{3}x\, \sqrt{g} \, G_{n}(x, x) \ ,
\end{equation}
where $G_{n}(x,x)$ is the coincident point Green's function on $\CM_n$.

Using the conformal transformation \eqref{S^1timesH^2} and the relation between the free energies \eqref{FreeEnergyRelation}, 
it is equivalent to that on $\BS^1_n\times \BH^2$,
\begin{equation}\label{dFntildedm2}
\frac{\partial}{\partial m^{2}} F_{n} = \frac{1}{2} \int_{\BS^1_n\times \BH^2} d^{3}\tilde x \sqrt{\tilde g} \ e^{-2\sigma(\tilde x)}\tilde G_{n}(\tilde x, \tilde x) \ ,
\end{equation}
with the dilaton field
\begin{align}
        e^{-2\sigma(\tilde x)} = \frac{\sin^2 \theta_0}{(\cos\tau+ \cos\theta_0\,\cosh u)^2 + \sin^2\theta_0 \sinh^2u } \ .
\end{align}
There appears the coincident point Green's function $\tilde G_n (\tilde x, \tilde x)$  in \eqref{dFntildedm2} which is independent of the position $\tilde x$ due to the homogeneity of $\BS^1_n\times \BH^2$.
This comes out of the integral and we are left with the integral of the conformal factor. 
There is a UV divergence in the coincident point Green's function which is canceled in the combination $ F_n - n\, F_1$:
\begin{align}
        \frac{\partial}{\partial m^2} \left( F_n - n\,  F_1\right) = \frac{V_n}{2} \left[\tilde G_{n}(\tilde x, \tilde x)|_{m^{2}=0}- n\,\tilde G_{1}(\tilde x, \tilde x)|_{m^{2}=0}\right] + O(m^{2}) \ ,
\end{align}
where $V_n$ is the integral of the conformal factor on $\BS^1_n\times \BH^2$,\footnote{The detail of the integral \eqref{VolumeIdentity} can be found in appendix \ref{ss:integral}. }
\begin{align}\label{VolumeIdentity}
        V_n = \int_{\BS^1_n\times \BH^2} d^{3}\tilde x \sqrt{\tilde g} \, e^{-2\sigma(\tilde x)} = 2 n\, \pi^3 \sin\theta_0\, R^3 \ .
\end{align}

There remains the coincident point Green's function which can be obtained by constructing the eigenfunctions of the scalar field on $\BS^1_n\times \BH^2$ (see e.g. \cite{Klebanov:2011uf}).
Inspecting the results in \cite{Klebanov:2011uf,Klebanov:2012va} we find
\begin{align}
        \lim_{n\to 1} \partial_n \left[\tilde G_{n}(\tilde x, \tilde x)|_{m^{2}=0}-n\, \tilde G_{1}(\tilde x, \tilde x)|_{m^{2}=0}\right] = - \frac{1}{32R} \ .
\end{align}
Finally, the replica trick \eqref{ReplicaTrick} yields the leading behavior of the EE of the cap-like region in the small mass limit:
\begin{align}\label{SmallMassExp}
 S_A(\theta_0, mR)  = \alpha\, \frac{2\pi R\sin\theta_0}{\epsilon} - F_\text{scalar} - \frac{\pi^3}{32} \sin\theta_0\, (mR)^2 + O\left( (mR)^4\right) \ ,
\end{align}
with $F_\text{scalar} = - (\ln 2)/8 + 3\zeta(3)/(16\pi^2) \approx 0.0638$ \cite{Klebanov:2011gs}. 

It is easy to see from the above expression that the REE $\CF_\text{LM} (R,\theta_0)$, as defined in \eqref{RenEE1}, is not stationary at the UV-fixed point $mR=0$ for any value of $\theta_0$.
We will discuss it in detail in section \ref{ss:NumericalResults}.

\subsection{Large mass expansion}\label{ss:LargeMassExp}
Although the analytic calculation of entanglement entropy is intractable even for free field theories if not conformal, one can expect to find a systematic expansion of the entropy for theories with a large mass gap $m$ in powers of $1/m$ \cite{Grover:2011fa,Klebanov:2012yf}:
\begin{align}\label{EE_Gap}
        S_\Sigma = \alpha \frac{\ell_\Sigma}{\epsilon} + \beta\,m\,\ell_\Sigma - \gamma_\Sigma + \sum_{n=0}^\infty \frac{c^\Sigma_{2n+1}}{m^{2n+1}} \ .
\end{align}
Here $\gamma_\Sigma$ is the topological entanglement entropy \cite{Kitaev:2005dm,levin2006detecting} depends on only the topology of the entangling surface $\Sigma=\partial A$ and $\beta$ is a scheme-independent constant \cite{Hertzberg:2010uv,Lewkowycz:2012qr}, while the numerical constant $\alpha$ is scheme-dependent.
For example, a free massive scalar field has $\beta=-1/12$ and $\gamma_\Sigma=0$.
The coefficients $c^\Sigma_{2n+1}$ are to be given by local integrals of functions of the extrinsic curvatures and its derivatives on the entangling surface because of the short-range correlation of order $1/m$ near the surface.
We note that the expansion \eqref{EE_Gap} has no proof for its validity in general, but is likely to hold for any entangling surface that is the disjoint union of a set of smooth curves without self-intersections \cite{Klebanov:2012yf,Nakaguchi:2014pha}.

Turning to the coefficients $c^\Sigma_{2n+1}$ for a free field theory, one can systematically determine $c^\Sigma_{2n+1}$ by the coefficient of the logarithmically divergent term of the entanglement entropy in the $(2n+4)$-dimensional massless free field theory compactified on $\BT^{2n+1}$ \cite{Huerta:2011qi,Klebanov:2012yf}.
The compactification yields an infinite tower of massive fields in $(2+1)$ dimensions, and the entanglement entropy across the entangling surface $\Sigma_{2n+2} = \Sigma \times \BT^{2n+1}$ should equal to the sum of the entropies for the massive fields across $\Sigma$.
One finds that the sum of the entropy of the order $1/m^{2n+1}$ terms over the Kaluza-Klein modes gives rise to a logarithmic UV divergence, which should be equated with the conformal anomaly term $S^{(2n+4)}_{\Sigma_{2n+2}}\big|_\text{log} = s^{(2n+4)}_{\Sigma_{2n+2}}\log \epsilon$ in the higher-dimensional theory.
Inspections of these coefficients lead to the following relation \cite{Klebanov:2012yf}
\begin{align}\label{KNPS}
        c^\Sigma_{2n+1} = - \frac{\pi (2\pi)^n(2n-1)!!}{\text{Vol}(\BT^{2n+1})} s^{(2n+4)}_{\Sigma_{2n+2}} \ ,
\end{align}
which can be used to determine the coefficients $c^\Sigma_{2n+1}$ from the conformal anomaly term in $(2n+4)$ dimensions.

To work it out explicitly for $c^\Sigma_1$ (i.e., $n=0$), we start with a four-dimensional theory wrapped on $\BS^1$ and consider an entangling surface $\Sigma_2 = \Sigma\times \BS^1$ wrapped on the $\BS^1$.
The entanglement entropy has a logarithmic divergence $S^{(4)}_{\Sigma_2}\big|_\text{log} = s^{(3+1)}_{\Sigma_2} \log \epsilon$ whose coefficient is known as Solodukhin's formula \cite{Solodukhin:2008dh,Fursaev:2013fta}:
\begin{align}\label{SolodukhinFormula}
        s^{(3+1)}_{\Sigma_2} = \frac{a}{2}\chi[\Sigma_2] + \frac{c}{2\pi} \int_{\Sigma_2} \left(\CR_{aa}  - \CR_{abab} - \frac{\CR}{3} + k^a_{\mu\nu}k_a^{\mu\nu} - \frac{1}{2}(k^{a\,\mu}_{\mu})^2\right) \ ,
\end{align}
where $\chi[\Sigma_2]$ the Euler characteristic, $\CR$ the Ricci scalar, $\CR_{aa} = \CR^{\mu\nu} n^a_\mu n^a_\nu$, $\CR_{abab} = \CR^{\mu\nu\rho\sigma} n^a_\mu n^b_\nu n^a_\rho n^b_\sigma$, and $k^a_{\mu\nu} = \gamma_\mu^\rho\gamma_\nu^\sigma \nabla_\rho n^a_\sigma$ is the extrinsic curvature for the normal vectors $n^a_\mu~(a=1,2)$ on $\Sigma_2$ with the induced metric $\gamma_{\mu\nu} = g_{\mu\nu} - n^a_\mu n^a_\nu$.
The central charges $a$ and $c$ are normalized so that a real scalar field has $(a,c) = \left(\frac{1}{180}, \frac{1}{120}\right)$.
In the present case where $\Sigma$ is topologically a circle, the entangling surface $\Sigma_2$ is topologically a torus with $\chi[\Sigma_2] = 0$, and only the second term \eqref{SolodukhinFormula} remains.
For a circular $\Sigma$ parametrized by $\theta = \Theta(\phi)$,
the timelike and spacelike unit normal vectors to $\Sigma_2$ are 
\begin{align}
        n^1_\mu = (1,0,0,0) \ , \qquad n^2_\mu = \frac{R \sin \theta}{\sqrt{\sin^2\theta + \left(\Theta'(\phi)\right)^2 }} (0,1,-\Theta'(\phi), 0) \ ,
\end{align}
where the fourth components are in the $\BS^1$ direction on which $\Sigma_2$ is wrapped.
The extrinsic curvature for the timelike normal vector $n^1$ vanishes due to the time translation invariance.
A short calculation shows that $\CR=2/R^2$, $\CR_{aa}=1/R^2$, $\CR_{abab}=0$, and
\begin{align}
        \kappa^2 \equiv k^2_{\mu\nu}k_2^{\mu\nu} - \frac{1}{2}(k^{2\,\mu}_{\mu})^2 =  \frac{\left( 2\cos\Theta\, \Theta'^2 + \sin\Theta  (\sin\Theta \cos\Theta - \Theta'') \right)^2}{2R^2\left(\sin^2\Theta + \Theta'^2 \right)^3} \ .
\end{align}
Combining with \eqref{KNPS} and \eqref{SolodukhinFormula}, we find
\begin{align}\label{c_1Sigma}
        c_1^\Sigma = - \frac{c}{2} \int_\Sigma \left[ \frac{1}{3R^2} + \kappa^2 \right] \ .
\end{align}
A few comments are in order:
\begin{itemize}
\item
Our result \eqref{c_1Sigma} for the coefficient $c_1^\Sigma$ reproduces that of \cite{Klebanov:2012yf} in the $R\to \infty$ and $\Theta\to 0$ limit with $R\sin\Theta$ kept fixed, under which the entangling surface becomes a curve on $\BR^2$.

\item 
We assumed that $\Sigma$ is a single curve on a sphere so far, but this result \eqref{c_1Sigma} holds for any entangling surface which is a disjoint union of curves because the uplifted entangling surface $\Sigma_2$ in $(3+1)$ dimensions is a disjoint union of tori whose Euler characteristics vanish and the integral \eqref{c_1Sigma} over $\Sigma$ is just the sum of the integrals over all disjoint curves.
\end{itemize}

In particular, a cap-like entangling region with opening angle $\theta_0$ is defined by $\Theta(\phi) = \theta_0$, and the coefficient \eqref{c_1Sigma} takes the simple form:
\begin{align}
        c_1^\text{cap} = - \frac{c\,\pi}{R}\sin\theta_0 \left[ \frac{1}{3} +\frac{\cot^2\theta_0}{2}\right] \ .
\end{align}
Combined with the expansion \eqref{EE_Gap} we find
\begin{align}\label{LargeMassExp}
  S_A (\theta_0, mR) = \alpha\, \frac{2\pi R \sin\theta_0}{\epsilon} -
  \frac{\pi}{6}mR\sin\theta_0
  - \frac{\pi}{120\, mR\sin\theta_0}\left(\frac{1}{2}
  - \frac{\sin^2\theta_0}{6}\right)
  + O\left((mR)^{-3}\right) \ ,
\end{align}
in the large $mR$ limit.

\subsection{Numerical results}\label{ss:NumericalResults}
We numerically calculated the EE of the cap-like region $A$
on the cylinder $\mathbb{R}\times\mathbb{S}^2$
by putting a free massive scalar field on the
lattice. 
We closely follow the method of \cite{Srednicki:1993im,Huerta:2011qi} whose details are found in appendix \ref{ss:Numerics}.

\begin{figure}[htbp]
\begin{center}
  \includegraphics[width=12cm]{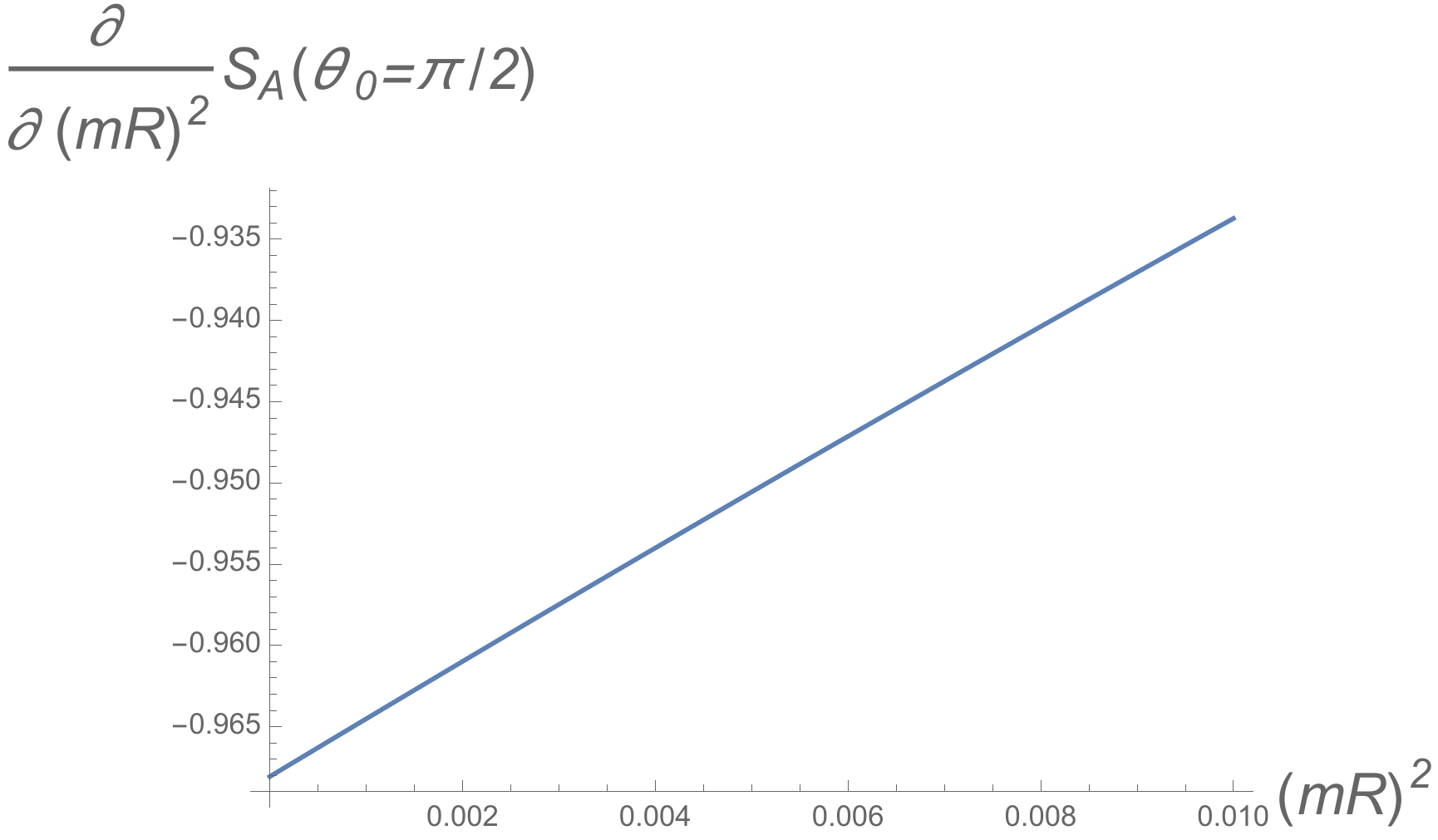}
\end{center}
\caption{
  The $(mR)^2$ derivative of the bare entanglement entropy $S_A(\theta_0=\pi/2)$
  of the hemisphere $A$,
  in the small mass region $mR \ll 1$.
  In taking the $(mR)^2$ derivative, we calculate bare entropies $S_A(\theta_0=\pi/2)$
  increasing $(mR)^2$, and fit it as a function of $(mR)^2$. 
  The lattice size is taken as $N=1001$. 
}
\label{fig:del_uv}
\end{figure}

Firstly, we check if the expansion \eqref{SmallMassExp} derived by using the conformal decompactification is valid in the small mass limit.
We calculate the derivative of $S_A(mR, \theta_0=\pi/2)$ with respect to $(mR)^2$ to avoid the UV divergence which contaminates our numerical precision.
Fig.\,\ref{fig:del_uv} shows that 
the entropy has a linear slope
\begin{align}
  \frac{\partial}{\partial (mR)^2}S_A(mR, \theta_0=\pi/2)
  = -0.968 + O\left((mR)^2\right) \ ,
\end{align}
in the small mass region $mR \ll 1$.
Integrating it by $(mR)^2$ leads that the entropy takes the form of
\begin{align}
  S_A(mR, \theta_0=\pi/2) = S_A(0, \theta_0=\pi/2) - 0.968\,(mR)^2 + O\left((mR)^4\right)\ .
\end{align}
Reassuringly, this is consistent with the analytic expression \eqref{SmallMassExp} for $\theta_0 = \pi/2$ with $-\pi^3/32 \approx -0.969$.

\begin{figure}[htbp]
  \includegraphics[width=8cm]{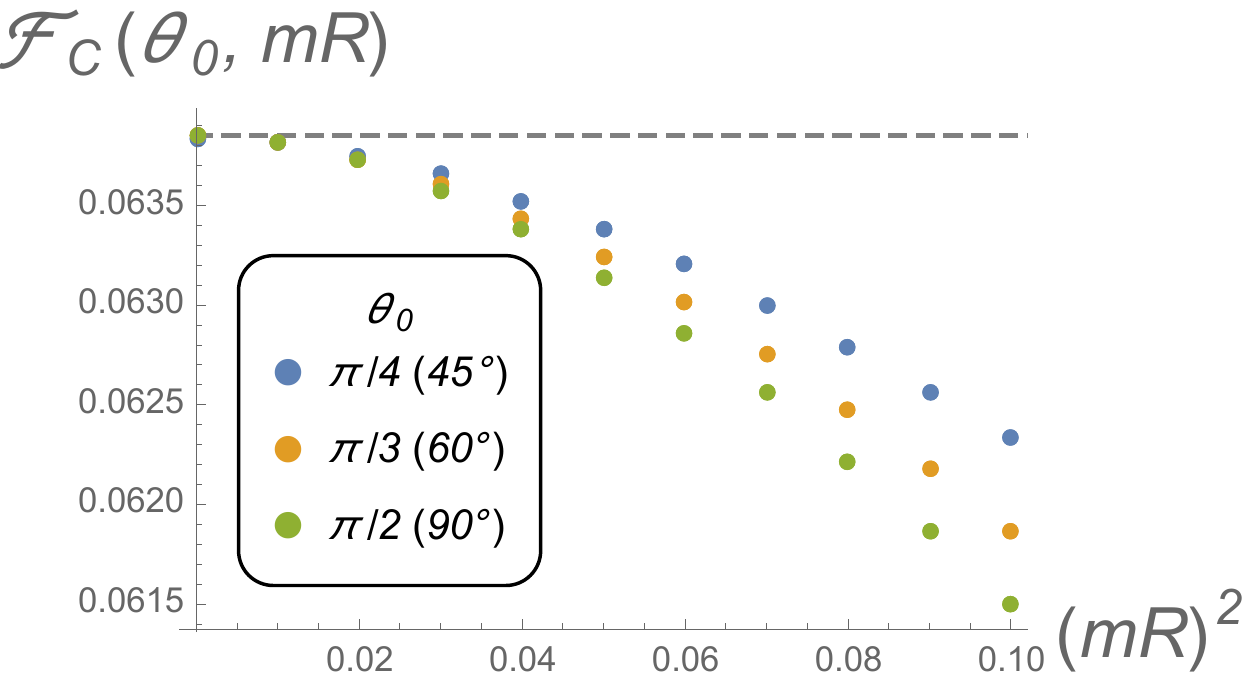}
  \includegraphics[width=8cm]{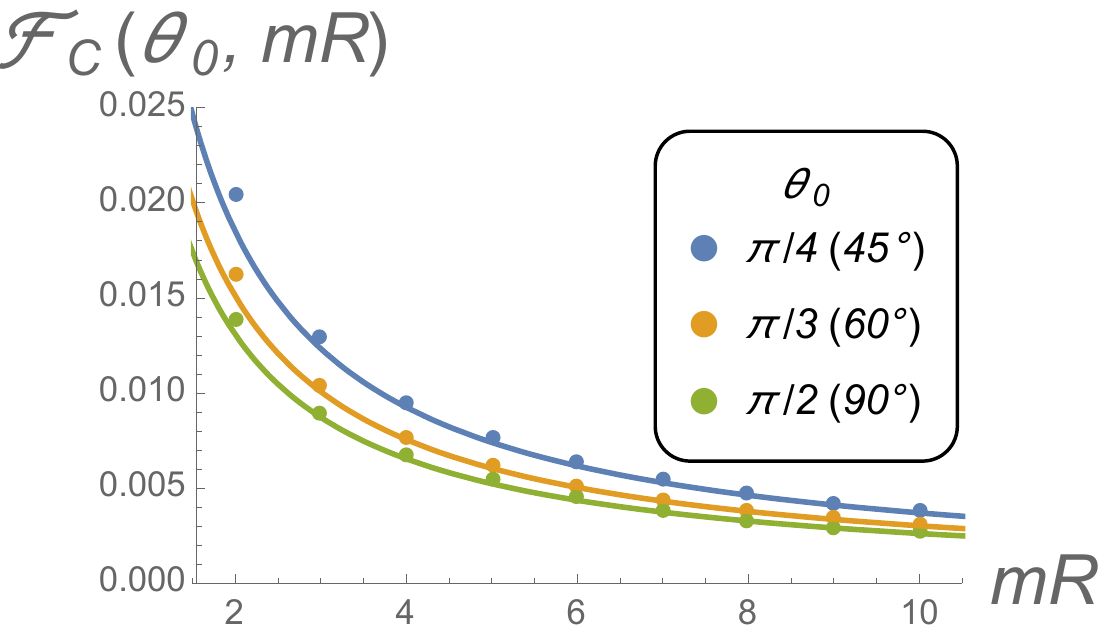}
\begin{center}
   ($a$)\hspace{7cm} ($b$)
\end{center}
\caption{
 The $mR$ dependence of the REE $\CF_\text{C}(mR,\theta_0)$
 in the small mass region $mR\ll 1$ ($a$)
 and the large mass region $mR\gg 1$ ($b$),
 with different cap angles $\theta_0=\pi/4$ (blue dots), $\pi/3$ (yellow dots) and $\pi/2$ (green dots).
The lattice size is taken as $N=501$.
 ($a$) The $\CF_\text{C}$ starts from a value $\CF_\text{C}(mR=0)\simeq0.06385$ (gray dotted line) at $mR=0$ with vanishing slope with respect to $(mR)^2$. This result reproduces the expected small mass expansion \eqref{c1cap}
 $\CF_\text{C} =F_\text{scalar}+O\left((mR)^4\right)$ in the small mass region $mR\ll 1$, which means that $\CF_\text{C}$ starts from the UV CFT value $F_\text{scalar}\simeq0.06381$ at $mR=0$ without any first order term of $(mR)^2$.
 ($b$) It asymptotes to the leading term $\pi/(120 mR\sin\theta_0)$ (solid lines) of the expected large mass expansion \eqref{LargeMassCurve}
 $\CF_\text{C} = \pi/(120 mR\sin\theta_0) +O\left(1/(mR)^3\right)$
 in the large mass region $mR\gg1$.
}
\label{fig:fc_uvir}
\end{figure}

Next we examine the REE $\CF_\text{C}(mR,\theta_0)$ defined by \eqref{RenEE2} to inspect the dependences of the entropy on $mR$ and $\theta_0$ in the broader ranges.
A detailed plot in the small mass region is shown in Fig.\,\ref{fig:fc_uvir} ($a$).
The REE starts decreasing from $0.0638 \approx F_\text{scalar}$ at the UV fixed point $mR=0$ for any $\theta_0$.
Furthermore, it is stationary in the sense that the first derivative with respect to $(mR)^2$ vanishes at the UV fixed point:
\begin{align}\label{c1cap}
  \CF_\text{C} (mR, \theta_0)=F_\text{scalar} +O\left( (mR)^4 \right)\ ,
\end{align} 
as predicted by the small mass expansion \eqref{SmallMassExp}.
This is in contrast to the REE $\CF$ of a disk \cite{Liu:2012eea} which is not stationary at the UV fixed point of a free massive scalar theory \cite{Klebanov:2012va,Nishioka:2014kpa} though our $\CF_C$ is supposed to reduce to $\CF$ in the flat space limit.
This difference may stem from the existence of the IR divergence on the flat space, which is regularized by the size of the sphere in the present setup.

In the other extreme limit of the large $mR$ region, we find $\CF_\text{C}$ decays to zero monotonically as shown in Fig.\,\ref{fig:fc_uvir} ($b$).
Comparing with the expansion \eqref{LargeMassExp} which yields the large mass behavior of the REE
\begin{align}\label{LargeMassCurve}
  \CF_\text{C} (mR,\theta_0) = \frac{\pi}{120\, mR\sin\theta_0} + O\left( (mR)^{-3}\right) \ ,
\end{align} 
our numerical data are well-fitted by the curves given by \eqref{LargeMassCurve} as $mR$ becomes large.

\begin{figure}[htbp]
\begin{center}
  \includegraphics[width=10cm]{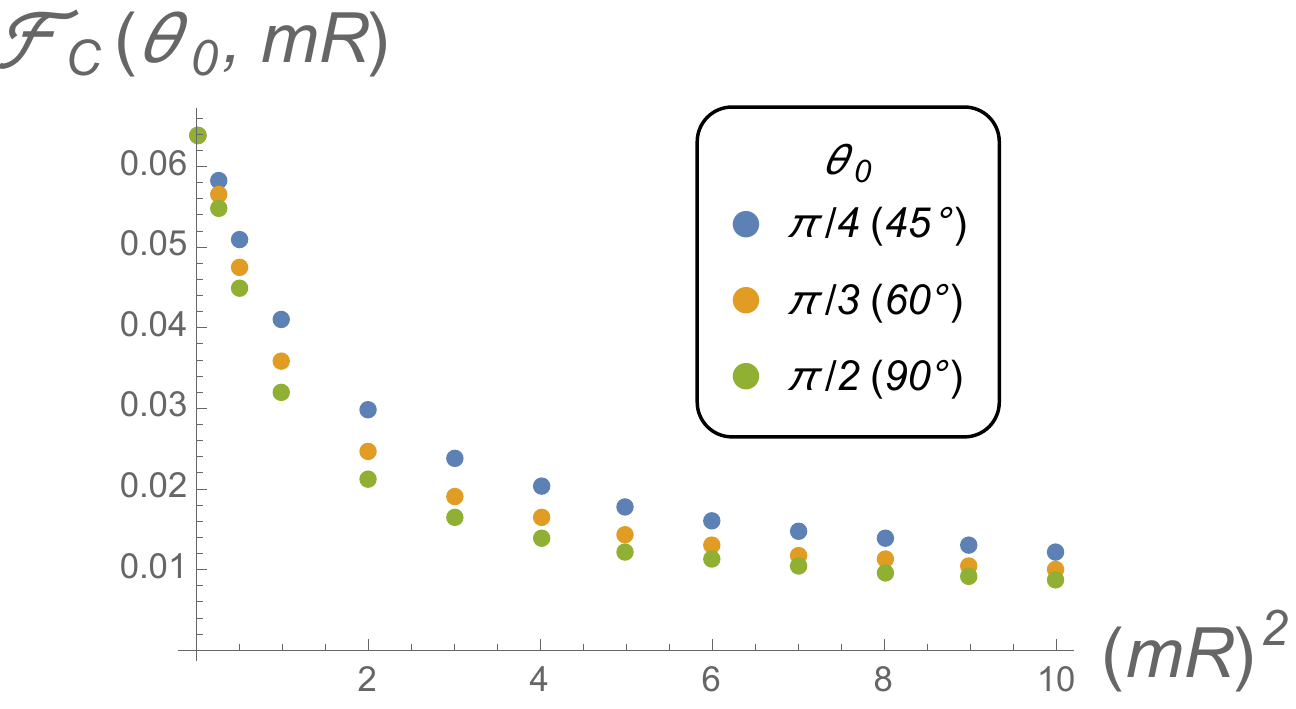}
\end{center}
\caption{The $mR$ dependence of the REE
 $\CF_\text{C}(mR,\theta_0)$
 with different cap angles $\theta_0=\pi/4$ (blue dots), $\pi/3$ (yellow dots) and $\pi/2$ (green dots). It is monotonically decreasing for the all $mR$.
 In taking the $\theta_0$ derivative, we calculate bare entropies $S_A(\theta_0)$
  increasing $\theta_0$, and fit it as a function of $\theta_0$. 
  The lattice size is taken as $N=501$.
}
\label{fig:fc_all}
\end{figure}

The whole shapes of the REEs are depicted in Fig.\,\ref{fig:fc_all}.
Clearly, the REEs are finite and monotonically decreasing to zero as $mR$ is increased for any $\theta_0$.
Also it is a monotonic function of $\theta_0$ for fixed $mR$, implying that increasing $\theta_0$ from $0$ to $\pi/2$ can be regarded as an RG flow.
The behavior of $\CF_\text{C}(mR, \theta_0)$
is reminiscent of the REE of a disk on the flat space \cite{Liu:2012eea,Klebanov:2012va,Nishioka:2014kpa} and the proof of monotonicity might proceed along the same lines as the proof of the $F$-theorem in \cite{Casini:2012ei}.

To recapitulate, all of the numerical results correctly reproduce
both the small and large mass expansions
in the previous two subsections,
and shows that the REE $\CF_\text{C}$ \eqref{RenEE2}
always decreases monotonically
with both the scale $mR$ and the cap size $\theta_0$ increased.
These numerical calculations give non-trivial checks
for the conformal decompactification method introduced in section \ref{ss:ConfDecom} and the Solodukhin's formula \eqref{SolodukhinFormula} on a curved space
because our numerical algorithm does not rely on any replica trick.

\section{Comments on $\CF_\text{LM}$}\label{ss:discussion}

In section \ref{ss:REECylinder}, we introduced two types of REEs on the cylinder which lead to the REE of a disk \eqref{LMREEoriginal} in the flat space limit, and we have solely dealt with the analytic and numerical properties of $\CF_\text{C}$ defined by \eqref{RenEE2} so far.
Here we will examine the Liu-Metzei type REE \eqref{RenEE1} using a free massive scalar field for comparison.
Contrary to the agreements for $\CF_\text{C}$, as we will see soon below, there appears an incongruity between the large mass expansion and the numerical calculation.
We will discuss possible resolutions to this puzzling situation in the end of this section.

\subsection{Analytic results in small and large mass limits}

The small mass expansion \eqref{SmallMassExp} of the EE of a free massive scalar field of mass $m$ leads to the small mass behavior of the Liu-Mezei type REE
\begin{align}\label{LM_smallexpansion}
    \CF_\text{LM}(mR, \theta_0) = F_\text{scalar} - \frac{\pi^3}{32}\sin\theta_0 (mR)^2 + O\left( (mR)^4\right) \ ,
\end{align}
which decreases linearly in $(mR)^2$ around the UV fixed point.
It is \emph{not} stationary in the Zamolodchikov's sense as the REE on the flat space is not \cite{Liu:2012eea,Klebanov:2012va,Nishioka:2014kpa} while  \eqref{LM_smallexpansion} is not a sensible expansion on the flat space because the $O(m^2)$ term diverges in the limit $R\to \infty$ and $\theta_0\to 0$ with $R\,\theta_0$ fixed.
It clearly shows that the breakdown of the perturbation theory emanates from the IR divergence, the volume of the flat space \cite{Herzog:2013py,Nishioka:2014kpa}.
The REE on the cylinder, on the other hand, regularizes both the UV and IR divergences and is suited to the perturbative expansions.

Similarly using \eqref{LargeMassExp}, the leading term in the large mass expansion is given by
\begin{align}\label{LM_LargeMass}
    \CF_\text{LM}(mR, \theta_0) = \frac{\pi}{120\, mR\sin\theta_0}\left( 1 - \frac{\sin^2\theta_0}{3}\right) + O\left( (mR)^{-3}\right) \ ,
\end{align}
that equals to \eqref{LargeMassCurve} up to the $\theta_0$ dependent constant.
Thus it monotonically decreases as increasing $mR$ for fixed $\theta_0$ as well as increasing $\theta_0$ for fixed $mR$.
Both \eqref{LargeMassCurve} and \eqref{LM_LargeMass} precisely reduce to the flat space result in \cite{Klebanov:2012va} as expected.
Our analytic results \eqref{LM_smallexpansion} and \eqref{LM_LargeMass} show the monotonic decrease of the Liu-Mezei type REE $\CF_\text{LM}$ in the small and large mass limits.

\subsection{Numerical results}\label{ss:F_LM_numeric}
In the numerical calculation of $\CF_\text{LM}$,
one can no longer use the same algorithm as for $\CF_\text{C}$ due to two obstacles.
One is that the definition \eqref{RenEE1} includes the derivative $\partial_R$ that requires the variation of the sphere radius $R$ as opposed to the previous case.
The other is that the discretization of the angle $\theta$ by
$\delta\theta=\pi/N$ causes the linear growth of the lattice spacing $\epsilon=R\delta\theta=\pi R/N$ in $R$, and 
one cannot remove by the differential operator $(R\partial_R-1)$ the $\epsilon$ dependence of the entanglement entropy
because the area law term $\alpha(2\pi R \sin\theta_0)/\epsilon=\alpha(2N\sin\theta_0)$ becomes independent of $R$.
To circumvent these obstacles, we employ a different regularization method; 
we calculate entropies by increasing both $mR$ and $N=\pi R/\epsilon$ simultaneously,
that is, fixing their ratio $mR/N=m\epsilon/\pi$,
and apply the differential operator $mR\,\partial_{(mR)}-1 = N\partial_N-1$ on the fitted results.
This prescription removes the dominant order $O(N)$ area law term successfully.
We checked that $\CF_\text{LM}(mR,\theta_0,N)$ becomes independent of $N$, namely, the REE $\CF_\text{LM}$ takes the same value for a different ratio $mR/N=m\epsilon/\pi$ as long as $mR$ is the same.
In this way, the $\epsilon$ dependence of the entropy is removed in the numerical calculation.

\begin{figure}[htbp]
\begin{center}
  \includegraphics[width=10cm]{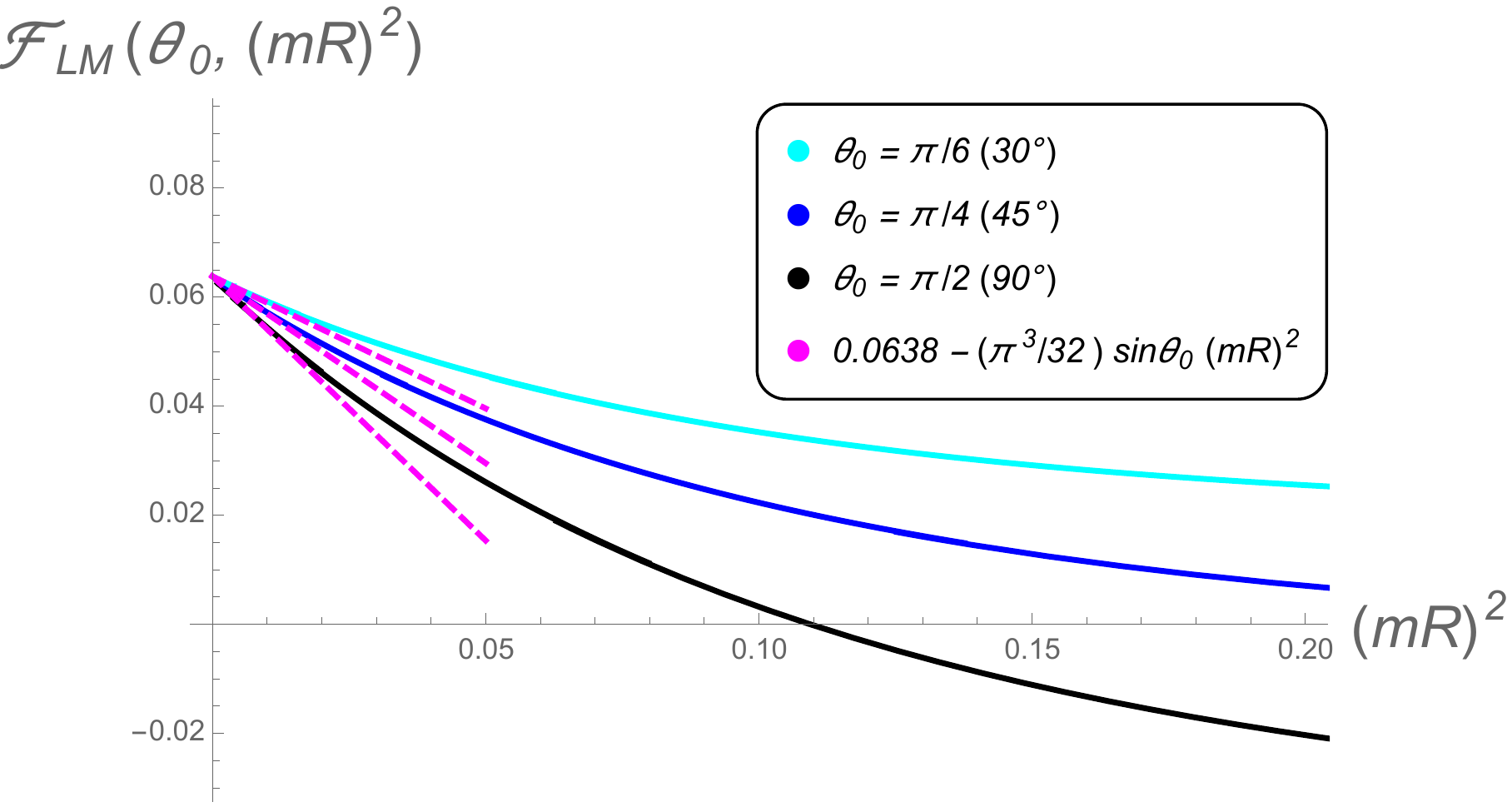}
\end{center}
\caption{The $(mR)^2$ dependence of the REE
 $\CF_\text{LM}(mR,\theta_0)$
 with different cap angles $\theta_0=\pi/6$ (light blue curve), $\pi/4$ (blue curve) and $\pi/2$ (black curve).
They correctly reproduce the small mass expansion (dotted magenta line)
in the small mass region $mR\ll1$.
 In taking the $R$ derivative, we calculate bare entropies $S_A(mR,N)$
  increasing both $mR$ and $N$ proportionally, and fit it as a function of $R$. 
}
\label{fig:fLM_uv}
\end{figure}

The resultant $\CF_\text{LM}$ correctly reproduces
the expected small mass expansion \eqref{LM_smallexpansion}
in the small mass region $mR\ll1$,
as shown in Fig.\,\ref{fig:fLM_uv}.
The REE $\CF_\text{LM}$ starts from the value
$\CF_\text{LM}|_{(mR)^2=0}=F_\text{scalar}\simeq0.0638$ of the UV CFT (a free massless scalar field)
like the REE $\CF_\text{C}$,
but in a non-stationary way $\partial_{(mR)^2}\CF_\text{LM}|_{(mR)^2=0}\neq 0$.

\begin{figure}[htbp]
\begin{center}
  \includegraphics[width=10cm]{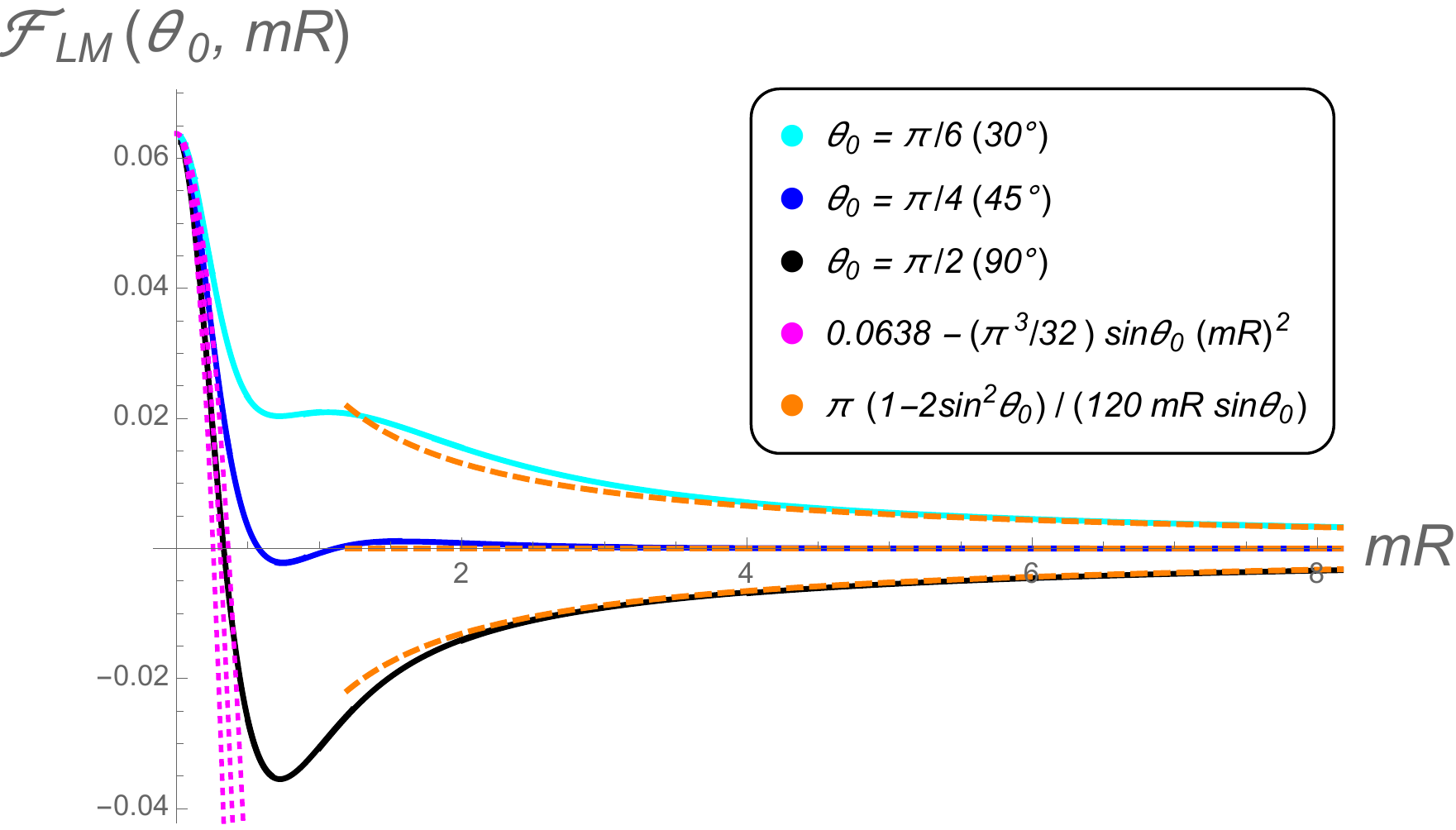}
\end{center}
\caption{The $mR$ dependence of the REE
 $\CF_\text{LM}(mR,\theta_0)$
 with different cap angles $\theta_0=\pi/6$ (light blue curve), $\pi/4$ (blue curve) and $\pi/2$ (black curve). They obey a large mass expansion $\CF_\text{LM}(mR, \theta_0) = \pi(
1 - 2\sin^2\theta_0)/(120\, mR\sin\theta_0) + O( (mR)^{-3})$ (orange dotted curve) different from the expected one \eqref{LM_LargeMass}.
}
\label{fig:fLM_all}
\end{figure}

We, however, unexpectedly found a discrepancy between the numerical result and the large mass expansion $\eqref{LM_LargeMass}$.
The whole $mR$ dependence of $\CF_\text{LM}$ is drawn in Fig.\,\ref{fig:fLM_all}.
In the large mass region $mR\gg1$,
The plot shows $\CF_\text{LM}$ asymptotes to the trivial IR CFT value $\CF_\text{LM}=0$, but it does not monotonically decrease in the large mass region. 
In fact, the numerically obtained REE $\CF_\text{LM}$ obeys a different large mass expansion 
\begin{align}\label{LM_LargeMass_Numerical}
    \CF_\text{LM}(mR, \theta_0) = \frac{\pi}{120\, mR\sin\theta_0}\left(
1 - 2\sin^2\theta_0\right) + O\left( (mR)^{-3}\right) \ ,
\end{align}
instead of the expected large mass expansion \eqref{LM_LargeMass}, where the coefficient of $\sin^2\theta_0$ is 6 times as large as that of \eqref{LM_LargeMass}.
For this equation to be true, the entanglement entropy should take the form of
\begin{align}\label{LargeMassExpNumerical}
  S_A (\theta_0, mR)\big|_\text{numerical} = \alpha\, \frac{2\pi R \sin\theta_0}{\epsilon} -
  \frac{\pi}{6}mR\sin\theta_0
  - \frac{\pi}{120\, mR\sin\theta_0}\left(\frac{1}{2}
  - \sin^2\theta_0\right)
  + O\left((mR)^{-3}\right) \ ,
\end{align}
in the large $mR$ limit. This is bigger than our expectation \eqref{LargeMassExp} by the amount of 
\begin{align}\label{Discrepancy}
  \frac{5}{6}\times\frac{c\,\pi \sin\theta_0}{mR}\ .
\end{align}
This discrepancy would not affect $\CF_\text{C}$
because it would be removed
by the operator $(\tan\theta_0\partial_{\theta_0}-1)$ in the definition of $\CF_\text{C}$.
It would vanish in the flat space limit
$R\to\infty$ and $\theta_0\to0$ with $R\theta_0$ fixed as commented below \eqref{c_1Sigma},
so it would not affect the result for $\CF_\text{C}$ in this limit.
Both expressions \eqref{LargeMassExp} and \eqref{LargeMassExpNumerical} of $S_A$ are consistent with the symmetry
$S_A(\pi/2-\theta_0)=S_A(\theta_0)$. 

\subsection{Discussion}
We have encountered an interesting puzzle that the numerical calculation is not in agreement with the large mass expansion \eqref{LM_LargeMass} of $\CF_\text{LM}$, whose derivation is based on the large gap expansion \eqref{EE_Gap}, the dimensional reduction of the $(3+1)$-dimensional free massless scalar field and the Solodukhin's formula \eqref{SolodukhinFormula}.
On the other hand, our numerical calculation appears to be working correctly because
the $\epsilon$ dependence in $\CF_\text{LM}$ is completely removed and
it exactly reproduces the small mass expansion \eqref{LM_smallexpansion} with the correct $\theta_0$ dependence as shown in Fig.\,\ref{fig:fLM_uv}.

The numerical result, if correct, implies that there is a missing contribution to the entropy in our derivation in section \ref{ss:LargeMassExp}.
A conceivable source of the discrepancy would be the postulated large gap expansion \eqref{EE_Gap}, where one could have included logarithmic terms of the forms $\log (m\,\ell_\Sigma)$ and $\frac{1}{m}\log (m\,\ell_\Sigma)$.
One can show the existence of such terms induces a new kind of UV divergences of the forms $\frac{1}{\epsilon}\log\epsilon$ and $\log^2 \epsilon$ in the uplifted free massless scalar theory in four dimensions.
This yields a contradiction as such terms have not been observed in the free scalar theory.
Thus the modification of the large gap expansion \eqref{EE_Gap} is unlikely to remedy the situation.

Although there exist the derivations for the formula based on the replica trick \cite{Fursaev:2013fta,Camps:2013zua} and holography \cite{Hung:2011xb}, another possible resolution of the discrepancy would be that there is a missing term in the Solodukhin's formula \eqref{SolodukhinFormula}.
Such a term, if exists, could depend on the curvature of the spacetime $\CM$ or the derivative of the extrinsic curvature along the normal direction to the entangling surface to account for the difference between \eqref{LargeMassExp} and the conjectured form \eqref{LargeMassExpNumerical} in the large mass limit.
Since the additional terms break conformal invariance 
we are not sure whether such terms are allowed to enter the formula.
A similar problem was raised in a recent paper \cite{Astaneh:2014sma}, where a total derivative term of the form $\Box \CR$ in the trace anomaly is argued to change the coefficient of the term proportional to $k^a_{\mu\nu}k_a^{\mu\nu}$ to the formula \eqref{SolodukhinFormula}.
The total derivative term is scheme-dependent in the sense that it can be removed by a counter term $\CR^2$ to the Lagrangian \cite{Birrell:1982ix}, and is usually ignored.
While adding the total derivative does not solve the present puzzle, it leaves us a possibility that the numerical calculation uses a different regularization scheme from the one in the large mass expansion \eqref{LM_LargeMass}.
In other words, the lattice regularization employed in section \ref{ss:F_LM_numeric} might not respect the conformal invariance of the uplifted theory in four dimensions. 

To pin down the source of the discrepancy, it is desirable to investigate the dependence of the entropy on the background curvature, that remains well-understood yet. 
We hope to address this intriguing puzzle in future publication.



\acknowledgments  
SB would like to thank Nilay Kundu for initial collaboration and valuable discussion. He would also like to thank Djordje Radicevic for very helpful discussions and collaboration on related projects. 
YN and TN would like to thank R.\,Jinno for valuable discussions. 
The work of SB and YN was supported by World Premier International Research Center Initiative (WPI), MEXT, Japan.
The work of YN was also supported in part by JSPS Research Fellowship for Young
Scientists.
The work of TN was supported in part by JSPS Grant-in-Aid for Young Scientists (B) No.\,15K17628.

\appendix
\section{Conformal transformation of a massive scalar field coupled to dilaton}\label{ss:appA}
For a free massive scalar our prescription for the conformal decompactification can be verified easily. The action of the theory is given by \eqref{ScalarAction}.
Let us denote the replica space free energy by $F_{n}$. It is easier to first calculate the derivative of $F_{n}$ with respect to the mass parameter $m^{2}$, which is given by the formula \eqref{dFndm2}.
The Green's function satisfies the equation, 
\begin{equation}
\left(-\nabla^{2} + \frac{\CR(g)}{8}+ m^{2}\right) G_{n} (x, x')= \frac{\delta^{3}(x- x')}{\sqrt{g}} \ .
\end{equation}

Now suppose we go to the conformally related manifold, $\widetilde\CM$ with metric $\tilde g_{\mu\nu}$ given by
\begin{equation}
g_{\mu\nu} = e^{-2\sigma} \tilde g_{\mu\nu} \ .
\end{equation}
It is easy to check that if we make the following transformation,
\begin{equation}\label{ConfMapDilaton}
G_{n}(x, x') = e^{\frac{\sigma(\tilde x)}{2}} \tilde G_{n} (\tilde x,\tilde
x' ) e^{\frac{\sigma(\tilde x')}{2}} \ ,
\end{equation}
then the Green's function equations get transformed into the Green's function equation on $\widetilde \CM$, but with a dilaton turned on:
\begin{equation}
\left(-\tilde \nabla^{2} + \frac{\CR(\tilde g)}{8}+ m^{2} e^{-2\sigma (\tilde x)}\right ) \tilde G_{n} ( x, x')= \frac{\delta^{3}(\tilde x-\tilde x')}{\sqrt{\tilde g}} \ .
\end{equation}
This equation comes from the action
\begin{align}\label{TildeActionScalar}
\begin{aligned}
        \tilde I &= \frac{1}{2} \int d^{3}\tilde x\, \sqrt{\tilde g} \,  \left[\tilde g^{\mu\nu}\partial_\mu \phi \partial_\nu \phi + \frac{\CR(\tilde g)}{8} \phi^{2} +  m^2 e^{-2\sigma(\tilde x)}\phi^{2}\right] \ , \\
        & = \tilde I_\text{CFT} +\frac{1}{2} \int d^{3}\tilde x\,\sqrt{\tilde g} \,  m^{2} e^{-2\sigma(\tilde x)} \phi^{2} \ .
\end{aligned}
\end{align}
For this action one can also compute the free energy and we can again write the derivative of the free energy as in \eqref{dFntildedm2}.
Use the transformations from \eqref{ConfMapDilaton}, then it is easy to see that
\begin{equation}\label{FFrelation}
\frac{\partial}{\partial m^{2}} F_{n} = \frac{\partial}{\partial m^{2}} \tilde F_{n} \ .
\end{equation}

We can work on the conformally transformed manifold as well. 
The free energies are the same modulo some additive constant independent of the mass parameter. Now let us see what is the interpretation of the tilde action. 
On the conformally transformed manifold, the dilaton couples to the massive scalar as
\begin{equation}
\tilde I[\tilde h,\tau(x)] = \tilde I_\text{CFT} +  \frac{1}{2} \int d^{3}\tilde x\,\sqrt{\tilde g} \, m^{2} e^{-2\tau(\tilde x)} \phi^{2} \ .
\end{equation}
From the general arguments in the section 2, we saw that if we work on the conformally transformed manifold then a compensating background dilaton field has to be switched on which is given by, $\tau(\tilde x) = \sigma (\tilde x)$. 
With this dilaton field the above action transforms precisely into the action we got in \eqref{TildeActionScalar}.

\section{Details of numerical calculations}\label{ss:Numerics}
In this appendix, we summarize the numerical algorithm
for calculating the entanglement entropy of the cap $A$ on the cylinder $\BR\times \BS^2$ for a conformally coupled free massive scalar field.

\subsection{Angular decomposition}
The action is given by \eqref{ScalarAction} with the Ricci scalar $\mathcal{R}=2/R^2$.
We can regard this theory as a free massive scalar field theory
\begin{align}
        I &= - \frac{1}{2}\int_{\BR\times \BS^2} d^3 x \sqrt{-g} 
        \left[ 
                g^{\mu\nu} \partial_\mu \phi \partial_\nu \phi + m^2_\text{eff}\, \phi^2
        \right] \ ,
\end{align}
with an effective mass
$m^2_\text{eff} \equiv m^2 + \frac{1}{4R^2}$.
The Hamiltonian is given as
\begin{align}\label{eq:hamiltonian}
  H = \int_0^\pi d\theta \int_0^{2\pi} d\phi~
      \frac{\sin\theta}{2}
      \left(
        R^2\pi^2 + (\partial_\theta \phi)^2 +
        \frac{(\partial_\phi \phi)^2}{\sin^2\theta} +
        m^2_\text{eff}\phi^2
      \right)\,,
\end{align}
where the conjugate momentum $\pi=\partial_t \phi$
satisfies the canonical commutation relation
\begin{align}
\begin{aligned}
  {[} \phi(\theta,\phi), \pi(\theta',\phi') ]
  &=
    \frac{i}{\sqrt{g}}
    \delta(\theta-\theta')\delta(\phi-\phi') \ ,\\
  &=
    \frac{i}{R^2\sin\theta}
    \delta(\theta-\theta')\delta(\phi-\phi')
    \label{eq:commutation}\,.
\end{aligned}
\end{align}

The region $A=\{(\theta,\phi);\,0<\theta<\theta_0\}$
has the rotational symmetry in the $\phi$ direction,
which allows us to reduce the space dimension to only the $\theta$ direction
by a following angular decomposition
\begin{align}
\begin{aligned}
  \phi(\theta,\phi) &=
    \frac{1}{\sqrt{\pi R \sin\theta}}
    \left(
      \frac{\phi_0(\theta)}{\sqrt{2}} +
      \sum_{n=1}^\infty
        \left(
          \phi_{-n}(\theta)\sin n\phi +
          \phi_{n}(\theta)\cos n\phi
        \right)
    \right) \ ,\\
  \pi(\theta,\phi) &=
    \frac{1}{\sqrt{\pi R^3 \sin\theta}}
    \left(
      \frac{\pi_0(\theta)}{\sqrt{2}} +
      \sum_{n=1}^\infty
        \left(
          \pi_{-n}(\theta)\sin n\phi +
          \pi_{n}(\theta)\cos n\phi
        \right)
     \right)\,.
\end{aligned}
\end{align}
In this angular decomposition,
the Hamiltonian \eqref{eq:hamiltonian}
\begin{align} \label{eq:hamiltonian2}
  H = \frac{1}{2R}\sum_{n=-\infty}^{\infty} \int_{0}^\pi d\theta
    \left[
      \pi_n^2(\theta) +
      \left((m_\text{eff}R)^2+\frac{n^2}{\sin^2\theta}\right)\phi_n^2(\theta)
      + \left(
        \sqrt{\sin\theta}\,\partial_\theta\left(\frac{\phi_n(\theta)}{\sqrt{\sin\theta}}\right)
        \right)^2
    \right] \,,
\end{align}
and the commutation relation \eqref{eq:commutation} becomes
\begin{align} \label{eq:commutation2}
  [\phi_n(\theta), \pi_{n'}(\theta')] &=
    i\delta_{nn'}\delta(\theta-\theta')\,.
\end{align}

\subsection{Lattice discretization}
We follow the discretization procedure \cite{Sabella-Garnier:2014fda}.
The space coordinate $\theta$ is discretized
as $\theta_j=j\pi/N$ $(j=1,2,\dots,N-1)$
with dynamic variables
\begin{align}
  (\Phi_n^j, ~ \Pi_n^j) \equiv
  \left\{ \begin{aligned}
    &\bigg(
      \sqrt{\frac{2\pi}{N}} \phi_n(\theta_j),
      &\sqrt{\frac{\pi}{2N}} \pi_n(\theta_j)
    \bigg) \,,
    \qquad & (j = 1, N-1)\,, \\
    &\bigg(
      \sqrt{\frac{\pi}{N}} \phi_n(\theta_j),
      &\sqrt{\frac{\pi}{N}} \pi_n(\theta_j)
    \bigg) \,,
    \qquad & (j\neq 1, N-1)\,.
  \end{aligned} \right.
\end{align}
In this discretization procedure,
the Hamiltonian \eqref{eq:hamiltonian2}
\begin{align}
  H & = \frac{1}{2R} \sum_{n=-\infty}^{\infty}
    \left(
      \sum_{j=1}^{N-1}(\Pi_n^j)^2 +
      \sum_{i,j=1}^{N-1}\Phi_n^i K_{ij}^{(n)} \Phi_n^j
    \right) \,,
\end{align}
and the commutation relation \eqref{eq:commutation2} becomes
\begin{align}
  [\Phi_n^j, \Pi_{n'}^{j'}] = i\delta_{nn'}\delta_{jj'}\,,
\end{align}
where $K_{ij}^{(n)}$ is
an $(N-1)\times(N-1)$ real symmetric tridiagonal matrix
\begin{align}
  K_{jj}^{(n)} &=
  \left\{ \begin{aligned}
    \frac{N^2}{\pi^2}\frac{\sin\theta_{3/2}}{2\sin\theta_1} +
    \frac{1}{4}\left(
      (m_\text{eff}R)^2+\frac{n^2}{\sin^2\theta_1}
    \right) \,,
    \qquad & (j = 1, N-1) \,,\\
    \frac{N^2}{\pi^2}2\cos\frac{\pi}{2N} +
    \left(
      (m_\text{eff}R)^2+\frac{n^2}{\sin^2\theta_j}
    \right) \,,
    \qquad & (j\neq 1, N-1) \,,
  \end{aligned} \right. \\
  K_{j,j+1}^{(n)} = K_{j+1,j}^{(n)} &=
  \left\{ \begin{aligned}
    &-\frac{N^2}{\pi^2}\frac{\sin\theta_{3/2}}{\sqrt{2\sin\theta_1\sin\theta_2}} \,,
    \qquad & (j = 1, N-2) \,,\\
    &-\frac{N^2}{\pi^2}\frac{\sin\theta_{j+1/2}}{\sqrt{\sin\theta_j\sin\theta_{j+1}}} \,,
    \qquad & (j\neq 1, N-2)\,,
  \end{aligned} \right. \\
  K_{ij}^{(n)} &= 0 \,, \qquad  (\,|i-j|>1\,) \,,
\end{align}
with a $\mathbb{Z}_2$ symmetry $K_{N-i,N-j} = K_{ij}$
corresponding the parity symmetry $\theta\to\pi-\theta$.
This matrix $K^{(n)}$ is related to the correlation matrices
$X_{ij}^{(n)} = \braket{\Phi_n^i\Phi_n^j}$
and
$P_{ij}^{(n)} = \braket{\Pi_n^i\Pi_n^j}$
as $X^{(n)} = \frac{1}{2}(K^{(n)})^{-1/2}$ and $P = \frac{1}{2}(K^{(n)})^{1/2}$.
The size $\theta_0$ of the subsystem $A$ is chosen to be a half-integer in units of
the lattice spacing,
$\theta_0 = (r+1/2)/N$ with an integer $r$.
This choice corresponds to the free boundary condition in the continuum limit.
In our calculation, we take $N=O(10^{2-3})$, which is sufficiently large for our purpose.

The entanglement entropy of the disk $S(\theta_0)$ is obtained by using
$r \times r$ submatrices
$X_r^{(n)}\equiv(X_{ij}^{(n)})_{1\le i,j\le r}$ and $P_r^{(n)}\equiv(P_{ij}^{(n)})_{1\le i,j\le r}$
as
\begin{align}\label{Entropy}
  S(\theta_0) = S_0 + 2\sum_{n=1}^{\infty} S_n \,,
\end{align} 
where $S_n$ is the contribution from the $n$-th angular mode
\begin{align}\label{eq:Sn}
 S_n = \mathrm{tr}[
   (C_n + 1/2) \log(C_n + 1/2) - (C_n - 1/2) \log(C_n - 1/2)
 ]\,,
\end{align}
with $C_n = \sqrt{X_r^{(n)}P_r^{(n)}}$.
In the following,
we describe how to perform this infinite summation
over $n$ under controlled numerical errors.

\subsection{Large angular momentum}
In the large angular momentum limit $n\to \infty$,
the correlation matrices
$X^{(n)}=\frac{1}{2}(K^{(n)})^{-1/2}$ and
$P^{(n)}=\frac{1}{2}(K^{(n)})^{1/2}$ approach almost diagonal matrices \cite{Klebanov:2012yf},
for general symmetric tridiagonal matrices $K^{(n)}$ such as
\begin{align}
\begin{aligned}
  K_{jj}^{(n)} &= k(j)n^2 + h(j)  \,,\\
  K_{j,j+1}^{(n)} = K_{j+1,j}^{(n)} &= t(j) \,,\\
  K_{ij}^{(n)} &= 0 \qquad (\,|i-j|>2\,)\,.
\end{aligned}
\end{align}
The products of the submatrices $X_r^{(n)} P_r^{(n)}$ almost equal
to $1/4$ times unit matrix up to order $1/n^8$.
The nontrivial entries are
at the lower-right corners
\begin{align}
\begin{aligned}
(X_r^{(n)} P_r^{(n)})_{rr} 
  &= \frac{1}{4} + \frac{c(r)}{n^4} - \frac{c(r)b(r)}{n^6} + O(1/n^8) \,,\\
(X_r^{(n)} P_r^{(n)})_{r,r-1}
  &= O(1/n^6) \,,\\
(X_r^{(n)} P_r^{(n)})_{r-1,r}
  &= O(1/n^6)\,,
\end{aligned}
\end{align}
where
\begin{align}
\begin{aligned}
  c(r)&\equiv
      \frac{t(r)^2}
    {4
    \sqrt{
      k(r)k(r+1)
    }
    \left(
      \sqrt{k(r)} +
      \sqrt{k(r+1)}
    \right)^2
    }\,,\\
  &= \frac{N^4}{\pi^4}\frac{\sin^2\theta_{r+1/2}}{(1/\sin\theta_r+1/\sin\theta_{r+1})^2}\,,\\
 b(r) &\equiv
        \frac{ 
                \frac{h(r)}{\sqrt{k(r)}} + \frac{h(r+1)}{\sqrt{k(r+1)}}
        }{
                \sqrt{k(r)}+\sqrt{k(r+1)}
        }+
        \frac{h(r)}{2k(r)} + \frac{k(r+1)}{2k(r+1)}\,,\\
        &=\frac{1}{2}\left((m_\text{eff}R)^2+\frac{N^2}{\pi^2}2\cos\frac{\pi}{2N}\right)(\sin\theta_r+\sin\theta_{r+1})^2\,.
\end{aligned}
\end{align}
The eigenvalues of the matrix $C_n = \sqrt{X_r^{(n)} P_r^{(n)}}$
are $1/2+O(1/n^8)$, except one eigenvalue
\begin{align}
  \frac{1}{2} + \frac{c(r)}{n^4} - \frac{c(r)b(r)}{n^6} + O(1/n^8)\,.
\end{align}
Therefore, most of the eigenvalues do not contribute to the $n$-th entanglement
entropy \eqref{eq:Sn} up to order $1/n^8$ and 
we obtain
\begin{align}\label{LargeSn}
S_n &=
  \frac{c(r)}{n^4} \left(1-\log \frac{c(r)}{n^4} \right) +
  \frac{c(r)b(r)}{n^6} \log \frac{c(r)}{n^4} + O(1/n^8) \ .
\end{align}
This asymptotic formula is much faster than the direct calculation of \eqref{eq:Sn}.

We perform the matrix trace calculation \eqref{eq:Sn} for $n$ less than some
large angular momentum $n_*$, and use this asymptotic formula \eqref{LargeSn}
for $n\ge n_*$ as long as $S_n(=O(\log n/n^4))$ is larger than the machine
precision.
The other higher modes are ignored.
We can make the numerical error sufficiently small
by taking $n_*$ large enough.\footnote{
See the last paragraph in the appendix of \cite{Nakaguchi:2014pha}
for the detail to determine $n_*$.}

\section{Integration of the conformal factor} \label{ss:integral}
We can exactly perform the integral $V_n$ \eqref{VolumeIdentity}
of the conformal factor on $\BS^1_n\times \BH^2$
\begin{align}
\begin{aligned}
  V_n &= \int_0^{2n\pi} d\tau \int_0^\infty du \int_{\BS^1} d\Omega_1
    ~(R^3\sinh u)\, e^{-2\sigma(\tau,u)} \,, \\
  &= 2n\pi R^3 \int_0^{2\pi}d\tau\int_0^\infty du ~
    \frac{\sinh u \, \sin^2 \theta_0}{(\cos\tau+ \cos\theta_0\,\cosh
u)^2 + \sin^2\theta_0 \sinh^2u } \ , \label{Vn}
\end{aligned}
\end{align}
thanks to an integration formula
\begin{align}
  \int_0^{2\pi}\frac{d\tau}{(\cos\tau+c)^2+s^2}
  =-\frac{i\pi}{s}\left[
    \frac{1}{\sqrt{(c-is)^{2}-1}} -
    \frac{1}{\sqrt{(c+is)^{2}-1}}
  \right]\,,
\end{align}
which we will apply with $c=\cos\theta_0\cosh u$ and $s=\sin\theta_0\sinh
u$.
Because
$
  \cos\theta_0\cosh u\pm i\sin\theta_0\sinh u
  = \cosh(u\pm i\theta_0)
$, the integration formula tells us that
\begin{align}
\begin{aligned}
  \int_0^{2\pi}
    \frac{d\tau}
    {(\cos\tau+\cos\theta_0\cosh u)^2+\sin^2\theta_0\sinh^2 u}
  &=
  -\frac{i\pi}{\sin\theta_0\sinh u}\left[
    \frac{1}{\sinh(u-i\theta_0)} -
    \frac{1}{\sinh(u+i\theta_0)}
  \right] \,, \\
  &= \frac{2\pi\coth u}{\sinh^2 u + \sin^2\theta_0} \, .
\end{aligned}
\end{align}
Plugging this result into the original integration \eqref{Vn},
we finally obtain
\begin{align}
\begin{aligned}
  V_n
  &= 4 n \pi^2 \sin\theta^2_0 \, R^3
    \int_0^\infty
    \frac{d(\sinh u)}{\sinh^2 u + \sin^2\theta_0} \,,\\
  &= 2 n\pi^3 \sin\theta_0 \, R^3\,.
\end{aligned}
\end{align}

\bibliographystyle{JHEP}
\bibliography{EE_Cylinder}

\end{document}